\documentclass[12pt]{article}
\usepackage{a4}
\usepackage{amsfonts}
             \let\d=\delta
     \let\e=\epsilon
            \let\q=\theta       
 \let\k=\kappa
\let\l=\lambda                    \let\x=\xi
 \let\p=\pi

   \let\f=\phi
\let\c=\chi                      

\let\w=\omega

\let\X=\Xi

\let\Y=\Psi                   
\let\L=\Lambda

\let\la=\label

      \let\pa=\partial      \let\bm=\bibitem
\newcommand{\be}{\begin{equation}}
\newcommand{\ee}[1]{\label{#1}\end{equation}}
\newcommand{\bea}{\begin{eqnarray}}
\newcommand{\eea}{\end{eqnarray}}
\newcommand{\ra}{\rightarrow}

\newcommand{\dV}[1]{\frac{\d}{\d V(#1)}}
\newcommand{\delV}[2]{\frac{\d\, #1}{\d V(#2)}}
\newcommand{\dY}[1]{\frac{\d}{\d \Y (#1)}}
\newcommand{\delY}[2]{\frac{\d\, #1}{\d \Y (#2)}}
\newcommand{\dc}{\frac{\partial}{\partial c}}
\newcommand{\delc}[1]{\frac{\partial #1}{\partial c}}
\newcommand{\sect}[1]{\setcounter{equation}{0} \section{#1}}
\renewcommand{\theequation}{\thesection .\arabic{equation}}
\newcommand{\refer}[1]{(\ref{#1})}
\newcommand{\Blangle}{\Bigl \langle}
\newcommand{\Brangle}{\Bigr \rangle}
\newcommand{\cint}[2]{\oint_{#2} \, \frac{d #1}{4\p i}}
\newcommand{\XL}{( \X_1 - \L_0 )}

\newcommand{\mat}[1]{\mbox{$ #1 $}}

\begin{document}
\thispagestyle{empty}
{\hbox to\hsize{
\vbox{\noindent CPT-97/P.3483 \hfill May 1997  \\ NIKHEF 97-019\\
hep-th/9705114}}}
\noindent
\vskip2cm
\begin{center}

{\Large\bf The Chiral Supereigenvalue Model}
\vglue2cm

Gernot Akemann $^\dag$ and Jan C.\ Plef\/ka $^\ddag$
\vglue1cm
{$^\dag\, $ \it Centre de Physique Th\'eorique, CNRS}
\footnote{unit\'e propre de recherche 7061},\\
{\it Case 907 Campus de Luminy, 13288 Marseille, Cedex 9, France}\\
{\footnotesize akemann@cpt.univ-mrs.fr}\\
\vglue 1.0 cm
{$^\ddag\, $ \it NIKHEF}\\
{\it  P.O. Box 41882, 1009 DB Amsterdam, The Netherlands}\\
{\footnotesize plefka@nikhef.nl}

\vglue2cm
{\large  ABSTRACT}
\end{center}
A supereigenvalue model with purely positive bosonic eigenvalues is
presented and solved by considering its superloop
equations. This model represents the supersymmetric generalization
of the complex one matrix model, in analogy to the relation between
the supereigenvalue and the hermitian one matrix model. 
Closed expressions for all planar multi-superloop correlation functions 
are found. Moreover an iterative
scheme allows the calculation of higher genus contributions to the
free energy and to the correlators. Explicit results 
for genus one are given.

\noindent

\vfill
\setcounter{page}{0}
\pagebreak

\sect{Introduction}

The term supersymmetric matrix model by now is used for a number of
zero and one dimensional large $N$ supersymmetric theories. 
$SU(N)$ Super-Yang-Mills theories in the large $N$ limit
reduced to one time dimension are relevant for the description of 
supermembranes 
\cite{dWHN}, and recently this model has been proposed as a
non-perturbative formulation of M-theory \cite{Banks}. The zero dimensional
version of this model has been proposed to be connected to IIB strings
\cite{japannbi}. Scalar supersymmetric matrix models in zero dimensions
have been applied to the investigation of branched polymers and the
meander problem \cite{AMZ} and to $c=-2$ conformal field theories
coupled to 2d gravity \cite{klebwilkjandiss}, for a review see \cite{fermmm}.

However, following the successful application of random matrix models
to 2d quantum gravity and lower dimensional bosonic strings \cite{2dgrav}, 
there still is no supersymmetric {\it matrix} model at hand which achieves a 
description of discretized super--Riemann surfaces. 
Nevertheless, on the level of eigenvalue models a disrete approach to 
2d supergravity was established through the so-called supereigenvalue
model \cite{Alv1}, representing a supersymmetric generalization of
the hermitian one matrix model. 
The investigation of its double scaling limit \cite{Alv2} revealed the
relevance of the model for $N=1$ super--Liouville theory \cite{dict}.
Moreover in refs.\ \cite{Jan} a complete iterative solution of the model
was presented by one of the authors.

As the complex matrix model \cite{Mor} is just as well suited for the
description of 2d quantum gravity as the hermitian one, we wish to ask 
the question whether a supersymmetric generalization of the complex matrix
model exists. The complex model  is in some respect more simple than the 
hermitian model as it enjoys the following feature. In their pioneering
work \cite{AJM} Ambj\o rn, Jurkiewicz and Makeenko have been able to give
a closed universal expression for all planar correlation functions. Such
an expression has been found only for the complex model with a simple
one-arc support of the spectral density, given by a single parameter.
For the hermitian model \cite{AJM,Amb} or for a more complicated support 
\cite{GA,GA2} the multi--loop correlators remain universal, but have to 
be calculated successively. Furthermore a complete iterative 
solution for higher genera of the complex matrix model 
has been given in \cite{AKM}, where explicit results have been presented up to 
and including genus three.

The present work generalizes the complex matrix model
to a model we have called chiral supereigenvalue model, as it contains only
the positive half of the bosonic real eigenvalues. In section 2 the 
model is defined and its superloop equations are given. These equations 
are  solved in the planar
limit in section 3. Moreover closed expressions for all planar 
multi--superloop correlation functions are derived here. The full iterative
scheme for higher genera is presented in section 4 where explicit results are
given for genus one.

\sect{Superloop Equations}

The chiral supereigenvalue model is built out of a set of $N$ Grassmann odd
and even variables\footnote{$N$ is even.} $\q_i$ and $\l_i$. Its partition
function reads
\be
{\cal Z}=
(\prod_{i=1}^{N}\int_0^\infty d\l_i \,\int d\q_i ) \, \prod_{i<j}\, 
(\l_i - \l_j - \q_i\q_j ) \,\exp \Bigl ( - N
\sum_{i=1}^N \, [ V(\l_i) - \q_i \Y (\l_i)] \, \Bigr )
\ee{model}
where
\be
V(\l_i)= \sum_{k=0}^\infty g_k {\l_i}^k   \quad \mbox{and} \quad \,\,\,
\Y(\l_i) = \sum_{k=0}^\infty \x_{k+1/2}  {\l_i}^k,
\ee{VY}
the $g_k$ and $\x_{k+1/2}$
being Grassmann even and odd coupling constants, respectively.
Note that the bosonic integral over the $\l_i$ runs over ${\mathbb R}_+$.
One shows that this model obeys a set of super--Virasoro constraints
\be
{\cal L}_n\ {\cal Z}\ =\ 0 \ ,\ \ 
{\cal G}_{n+1/2}\ {\cal Z}\ =\ 0\ , \ \ n\geq 0\ ,
\ee{Vir}
where the operators ${\cal L}_n$ and ${\cal G}_{n+1/2}$ are given in 
\cite{Alv1}. The supereigenvalue model \cite{Alv1} obeys these constraints
starting already at $n\geq-1$. Hence the partition function of the
supereigenvalue model also solves the chiral constraints \refer{Vir}
but not vice versa.
A similar relation holds between the loop equations of the hermitian and
complex matrix model. 

We introduce the one--superloop correlators
\be
\widehat{W}(p \mid\, ) = N \,\Blangle \,\sum_i \frac{p\, \q_i}{p^2-\l_i}
 \,
\Brangle
\qquad
\mbox{and}
\quad
\widehat{W}(\, \mid p) = N \, \Blangle \,\sum_i \frac{p}{p^2-\l_i}\, 
\Brangle .
\ee{Wwidehat}
They may be obtained from
the partition function ${\cal Z}$ by application of the superloop insertion
operators $\d/ \d V(p)$ and  $\d/ \d \Y (p)$:
\be
\widehat{W}(p_1,\ldots ,p_n \mid q_1,\ldots ,q_m)= \frac{1}{\cal Z}
\dY{p_1} \,\ldots\,
\dY{p_n}\, \dV{q_1}\,\ldots\, \dV{q_m}
\,\, {\cal Z},
\ee{Wwidehatnm2}
where
\be
\dV{p}= -\sum_{k=0}^\infty\frac{1}{p^{2k+1}} \frac{\pa}{\pa g_k}
 \quad\mbox{and}\quad
\dY{p}= -\sum_{k=0}^\infty\frac{1}{p^{2k+1}} \frac{\pa}{\pa 
\x_{k+1/2}}.
\ee{superloopensertionops}
However, it is convenient to work with the connected part of the
$(n|m)$--superloop correlators, denoted by $W$. They may
be obtained in the following way from the free energy 
\mbox{$F=N^{-2}\,\ln\,{\cal Z}$}:
\bea
\lefteqn{W(p_1,\ldots ,p_n \mid q_1,\ldots ,q_m)=
\dY{p_1} \,\ldots\,
\dY{p_n}\, \dV{q_1}\,\ldots\, \dV{q_m}
\,\,  F}\label{Wnm}\\ & = &
N^{n+m-2} \,
\Blangle \,
\sum_{i_1} \frac{p_1\q_{i_1}}{p_1^2-\l_{i_1}} \, \ldots \, \sum_{i_n}
\frac{p_n\q_{i_n}}{p_n^2-\l_{i_n}} \, \sum_{j_1}\frac{q_1}{q_1^2-\l_{j_1}} \,
\ldots\, \sum_{j_m} \frac{q_m}{q_m^2-\l_{j_m}} \,\Brangle _C .
\nonumber\eea
Note that correlation functions with \mat{n\geq 2} 
vanish due to the structure of $F$ discussed below.
With the normalizations chosen above, one assumes that these correlators
enjoy the genus expansion 
\be
W(p_1,\ldots ,p_n \mid q_1,\ldots ,q_m)= \sum_{g=0}^\infty \, =
\frac{1}{N^{2g}}
\, W_g (p_1,\ldots ,p_n \mid q_1,\ldots ,q_m).
\ee{genusexpWnm}
Similarly one has the genus expansion
\be
F= \sum_{g=0}^\infty \, \frac{1}{N^{2g}} \, F_g
\ee{genusexpF}
for the free energy.

\subsection{Superloop Equations}

The superloop equations of our model are two Schwinger--Dyson equations.
For the Grassmann--odd superloop equation we perform the shift
\be
\l_i \ra \l_i + \q_i \frac{\e\, \l_i}{p^2-\l_i} \quad \mbox{and} \quad
\,\,\,
\q_i \ra \q_i + \frac{\e\,\l_i}{p^2-\l_i}
\ee{shift1}
where $\e$ is an odd constant. One then finds the odd equation
\be
\cint{\w}{C} \, \frac{\w\, \bar{V}^\prime\, (\w)W (\w \mid\, ) \, + \,
2\, \w^2\, \bar{\Y} (\w)\, W (\, \mid \w)}{p^2-\w^2} \ =\ 
W(p \mid \, ) \, W(\,\mid p) \, + \, \frac{1}{N^2} \, W(p \mid p)
\ee{superloop1}
and its counterpart, the Grassmann--even superloop equation, takes the form
$$
\cint{\w}{C} \, \frac{\w\,\bar{V}^\prime (\w ) \, W(\, \mid \w ) \, + 
\,\w\,
\bar{\Y}^\prime (\w ) \,
W(\w \mid \, )}{p^2 - \w^2} \, -\, \frac{p}{2} \, \frac{d}{dp}\, 
\cint{\w}{C} \,
\frac{\bar{\Y} (\w )\, W( \w \mid )\, }{p^2-\w^2} = 
\phantom{W(p \mid \, )\,}$$
 \be
 \frac{1}{2} \, \Bigl [ \, W(\, \mid p)^2 \, + \frac{1}{2p} \, 
W(p \mid \, )\, W^\prime (p
 \mid \, ) \, + \, \frac{1}{2\,N^2}\, \Bigl ( \, W(\, \mid p,p) \, - \,
\frac{1}{2p}\,\frac{d}{dq}\,
 W( q,p \mid \, ) \,  \Bigr | _{p=q} \, \Bigr ) \Bigr ] .
\ee{superloop2}
It is obtained through the shift
\be
\l_i \ra \l_i + \frac{\e\,\l_i}{p^2-\l_i} \quad \mbox{and} \quad \,\,\,
\q_i \ra \q_i + \frac{1}{2}\, \frac{\e\, p^2 \, \q_i}{(p^2-\l_i)^2 }
\ee{shift2}
with $\e$ even and infinitesimal. Moreover we have introduced the 
potentials
\bea
\bar{V}^\prime (\w) &=& 2\, \w\, V^\prime (\w^2) \nonumber\\
\bar{\Y}^\prime (\w) &=& 2\, \w\, \Y^\prime (\w^2) \nonumber\\
\bar{\Y} (\w) &=&  \Y (\w^2).
\eea
In the derivation we have assumed that the loop correlators have one--cut
structure,
i.e.\ in the limit $N\ra \infty$ we assume that the eigenvalues are
contained in a finite interval \mbox{$[-\sqrt{c},\sqrt{c}]$}.
Moreover $C$ is a curve around the cut.

The key to the solution of these complicated equations order by order
in $N^{-2}$ is the observation that
the free energy $F$ depends at most quadratically on fermionic coupling
constants. This was proven in ref.\ \cite{McA} for the supereigenvalue model,
but an inspection of the proof reveals that the same holds true for the
chiral supereigenvalue model as well.
Via eq. \refer{Wnm} this directly translates to the one-loop
correlators, which we from now on write as
\bea
W(p \mid \, ) &=& v(p) \\
W(\, \mid p)  &=& u(p) \, + \, \widehat{u}(p).
\eea
Here $v(p)$ is of order one in fermionic couplings, whereas $u(p)$ is taken
to be of order zero and $\widehat{u} (p)$ of order two in the fermionic
coupling
constants $\x_{k+1/2}$. This
observation allows us to split up the two superloop equations
\refer{superloop1}
and \refer{superloop2} into a set of four equations, sorted by their order
in the $\x_{k+1/2}$'s. Doing this we obtain

\medskip
\leftline{Order 0:}
\be
\cint{\w}{C} \, \frac{\w\, \bar{V}^\prime (\w )}{p^2-\w^2} \, u(\w ) = 
\frac{1}{2}\, u(p)^2
\, + \, \frac{1}{2\, N^2}\, \dV{p}\, u(p) \, + \, \frac{1}{4\, p}\,
\frac{1}{N^2}
\, \frac{d}{dq}\, \dY{p} \, v(q) \Bigr | _{p=q}
\ee{order0}
Order 1:
\be
\cint{\w}{C}\, \frac{\w\,\bar{V}^\prime (\w )}{p^2-\w^2 }\, v(\w )\, 
+ \, \cint{\w}{C} \,
\frac{2\,\w^2\, \bar{\Y} (\w )}{p^2-\w^2}\, u(\w ) = v(p)\, u(p) \, 
+ \, \frac{1}{N^2}\, \dV{p}\, v(p)
\ee{order1}
Order 2:
$$
\cint{\w}{C} \, \frac{\w\,\bar{V}^\prime (\w )}{p^2-\w^2 }\, =
\widehat{u}(\w )
 \,+\, 
\cint{\w}{C} \, \frac{\w\,\bar{\Y}^\prime (\w )}{p^2-\w^2}\, v(\w ) \, 
- \, \frac{p}{2}\,
\frac{d}{dp} \, \cint{\w}{C}\,  \frac{\bar{\Y} (\w )}{p^2-\w^2 }\, v(\w ) 
$$
\be
+ u(p)\, \widehat{u}(p)\, +\, \frac{1}{4p}\, v(p)\, \frac{d}{dp}\, v(p) \,
+ \,
\frac{1}{2\, N^2}\, \dV{p}\, \widehat{u}(p) 
\ee{order2}
Order 3:
\be
\cint{\w}{C}\, \frac{2\,\w^2\bar{\Y} (\w )}{p^2-\w^2 }\, \widehat{u} (\w ) 
= v(p)\, \widehat{u} (p).
\ee{order3}
Plugging the genus expansions into these equations lets them
decouple partially, in the sense that the equation of order 0 at genus $g$
only involves
$u_g$ and lower genera contributions. The order 1 equation then only
contains
$v_g$, $u_g$ and lower genera results and so on. The first thing to do,
however, is to find the solution for $g=0$.

\sect{The Planar Solution}

\subsection{Solution for $u_0(p)$ and $v_0(p)$}

In the limit $N\ra \infty$ the order 0 equation may be solved:
\be
u_0 (p)\, = \, \cint{\w}{C} \, \frac{\w\,\bar{V}^\prime (\w
)}{p^2-\w^2}\, 
\biggl [ \,
\frac{p^2-c}{\w^2-c}\, \biggr ]^{1/2} ,
\ee{u_0}
where the endpoint  $c$ of the cut on the positive real axis
is determined by the requirement:
\be
1=\cint{\w}{C}\, \frac{\w\, \bar{V}^\prime(\w )}{\sqrt{\w^2-c}},
\ee{determinexy}
deduced from our knowledge that $W(\, \mid p)= 1/p + {\cal O}(p^{-2})$.

The order 1 equation \refer{order1} in the $N\ra \infty $ limit determining
the odd loop correlator $v_0(p)$ may now also be solved to obtain

\be
v_0(p) \, =\, 2\, \cint{\w}{C}\,
\frac{\w^2\,\bar{\Y} (\w )}{p^2-\w^2}\, \biggl [ \, \frac{\w^2 -c}
{p^2-c} \, \biggr ]^{1/2} \,\, + \,\, \frac{\c}{\sqrt{p^2-c}} .
\ee{v_0}
Here $\c$ is a constant not determined by eq.\ \refer{order1}, in fact
$\c=N^{-1}\, \langle \, \sum_i \q_i \, \rangle$ in the planar limit.

\subsection{Moments and Basis Functions}

Instead of the couplings $g_k$
we introduce the bosonic moments $M_k$ and $I_k$ defined by 
\cite{AKM}\footnote{Note the different sign in $c$ chosen here.}
\bea
M_k &=& \cint{\w}{C}\, \frac{\w\,\bar{V}^\prime (\w )}{ \w^{2k+2}} \,
\frac{1}{[\, \w^2 -c \, ]^{1/2}}
, \quad k\geq 0 \\ &&\nonumber\\
I_k &=& \cint{\w}{C}\, \frac{\w\,\bar{V}^\prime (\w )}{ (\w^2 - c)^k} \,
\frac{1}{[\, \w^2 - c\, ]^{1/2}}
, \quad k\geq 0 ,
\eea
and the couplings $\x_{k+1/2}$ are replaced by the fermionic moments
\bea
\L_k &=& \cint{\w}{C}\,
\frac{\w^2\,\bar{\Y} (\w )}{w^{2k+2}}\, [\,\w^2 -c\, ]^{1/2}
,\quad k\geq 0 \nonumber \\ && \nonumber\\
\X_k &=& \cint{\w}{C}
\, \frac{\w^2\,\bar{\Y} (\w )}{(\w^2 -c)^k}\, [\,\w^2 -c\, ]^{1/2}
,\quad k\geq 0.
\eea
The main motivation for introducing these new variables is that, for each 
term
in the genus expansion of the free energy and the correlators, the
dependence on an
infinite number of coupling constants arranges itself nicely into a
function of a {\it finite} number of moments.

We further introduce the basis functions $\c ^{(n)}(p)$
and $\Y^{(n)}(p)$ recursively
\bea
\c ^{(n)}(p) &= &
\frac{1}{I_1}\, \Bigl ( \, \f^{(n)}(p)\, - \,\sum_{k=1}^{n-1} \c^{(k)}
(p)\, I_{n-k+1}\, \Bigr ) , \\
& &\nonumber \\
\Y^{(n)}(p) &= &\frac{1}{M_0}\,
\Bigl ( \, \Omega^{(n)}(p)\, - \,\sum_{k=1}^{n-1} \Y^{(k)}
(p)\, M_{n-k} \Bigr ) ,
\eea
where
\bea
\f^{(n)} (p) & = & \frac{1}{(p^2-c)^{n+1/2}}\, \la{bas1} \\
& & \nonumber \\
\Omega^{(n)}(p) & = & \frac{1}{p^{2n}\, (p^2-c)^{1/2}}, \la{bas2}
\eea
following again \cite{AKM}.
It is easy to show that for the linear operator 
$\widehat{\bar{V}}^\prime$ defined by
\be
\widehat{\bar{V}}^\prime \circ f(p) = 
\cint{\w}{C}\, \frac{\w\,\bar{V}^\prime (\w )}{p^2-\w^2}\,
f(\w )\, -\, u_0(p)\, f(p)
\ee{opVprime}
we have
\bea
\widehat{\bar{V}}^\prime\,\circ\, \c ^{(n)}(p) &=& \frac{1}{(p^2-c)
^n},  \label{VprimeChi} \\
& & \nonumber \\
\widehat{\bar{V}}^\prime \,\circ\, \Y^{(n)}(p) &=& \frac{1}{p^{2n}}. 
\label{VprimePsi}
\eea

\subsection{Solution for $\widehat{u}_0$ and $\c$}

Next consider the order 2 equation \refer{order2} at genus 0.
Plugging  eq.\
\refer{v_0} into the right hand side of this equation yields after a
somewhat lengthy calculation
\be
\widehat{\bar{V}^\prime}\,\circ\, \widehat{u}_0 \, =\,
\frac{\X_2\, (\, 2\,\X_1 - \c\, )}{p^2-c}
\ee{page20g}
With eqs.
\refer{VprimeChi} and \refer{VprimePsi} this immediately tells us that
\be
\widehat{u}_0(p) = 
\frac{\X_2\, (\, 2\,\X_1 - \c\, )}{I_1}\, \frac{1}{(p^2-c)^{3/2}}.
\ee{uhat0first}
Finally we determine the odd constant $\c$. One may think that this is
done by employing the \hbox{order 3} equation for $g=0$, however it
turns out that \refer{order3} is identically fulfilled with the
above results for $v_0(p)$ and $\widehat{u}_0(p)$. We instead use the
consistency requirement
\be
W_0(p,p\mid\, )= \dY{p}\, v_0(p) \, =\, 0
\ee{consist}
due to anti--commutativity. From this one has
\be
\dY{p}\,\c = p- \frac{\sqrt{p^2-c}}{2}-\frac{\sqrt{p^2-c}\, p^2}{2\, 
(p^2-c)},
\ee{getchi}
from which one deduces
\be
\c=\L_0\, +\, \X_1,
\ee{chi}
by using some of the formulas in the appendix. Putting it all
together, we may now write down the complete genus 0 solution for
the one--superloop correlators $W(\, \mid p)$ and $W(p\mid \,)$:

\bea
W_0(\,\mid p) &=&\cint{\w}{C}\, \frac{\w\,\bar{V}^\prime (\w
)}{p^2-\w^2}\,
\biggl [ \, \frac{p^2-c}{\w^2 -c}\, \biggr ]^{1/2}\,
+\,
\frac{\X_2\,(\, \X_1- \L_0\, )}{I_1\, (p^2-c)^{3/2}}
\nonumber\\
&&\nonumber\\&&\nonumber\\
W_0(p\mid \, ) &=& 2\, \cint{\w}{C}\, \frac{\w^2\, \bar{\Y} (\w
)}{p^2-\w^2}
\, \biggl [\,
\frac{\w^2 -c}{p^2-c}\, \biggr ] ^{1/2} \, +\, \frac{\X_1\, +\, \L_0}
{[\, p^2-c\, ]^{1/2}}.
\label{Wp0_0}
\eea

\subsection{All planar correlation functions}

According to the definition \refer{Wnm} all planar correlation functions 
can be obtained from eq.\ \refer{Wp0_0} by taking functional derivatives
w.r.t.\ $V(p)$ or $\Psi(p)$. In this subsection we prove that this successive
application can be reduced to a closed algebraic expression for all
multi-superloop correlators. 

Let us begin with the highest degree in fermions, the two--fermion loop
correlator, which takes the universal form
\be
W_0(p,q\mid\, )= \frac{1}{2(p^2-q^2)}\biggl [ p^2 \sqrt{\frac{q^2-c}{p^2-c}}
                 \ +\ q^2 \sqrt{\frac{p^2-c}{q^2-c}}\ -\ 2pq \biggl ] .
\ee{W0pq|}
>From comparing it 
to the well known two--loop correlator $W_0^{\mbox{\scriptsize cmm}}(p,q)$ 
of the complex matrix model \cite{AJM},
\be
W_0(p,q\mid\,)\ =\ 2(p^2-q^2)W_0^{\mbox{\scriptsize cmm}}(p,q)\ ,
\ee{W0rel}
a closed expression can be immediately written down following \cite{AJM}:
\be
W_0(p,q\mid r_1,\ldots,r_n)= \biggl ( \frac{1}{I_1}\dc\biggl )^{n-1}
  \frac{c(p^2-q^2)}{4I_1(p^2-c)^{3/2}(q^2-c)^{3/2}}
  \prod_{k=1}^n \frac{c}{(r_k^2-c)^{3/2}},\ \ n\geq 1.
\ee{W0pq|...}

For the one--fermion loop correlators we first calculate
\be
W_0(p\mid q)\ =\ \frac{1}{2I_1(p^2-c)^{3/2}(q^2-c)^{3/2}}
  \biggl [ (p^2-c)\, c\, \X_2 - p^2(\X_1-\L_0)\biggl ],
\ee{W0p|q}
using some formulas of the appendix. Due to the fact that the fermionic
moments $\X_k$ and $\L_k$ depend on the bosonic potential $V(p)$ only
via $c$, we can apply the following theorem proven in \cite{AJM}
\be
\dV{p}\ \biggl ( \frac{1}{I_1}\dc\biggl )^n \frac{1}{I_1} h(c)\ =\ 
     \biggl ( \frac{1}{I_1}\dc\biggl )^{n+1} \frac{1}{I_1} h(c)
        \frac{c}{(p^2-c)^{3/2}} \ ,\ \ n\geq 0\ ,
\ee{theo}
where $h(c)$ is an arbitrary function of $c$. This immediately leads 
to the closed form 
\be
W_0(p\mid q_1,\ldots,q_n)= 
 \biggl ( \frac{1}{I_1}\dc\biggl )^{n-1} 
 \frac{(p^2-c)\, c\, \X_2 - p^2(\X_1-\L_0)}{2cI_1(p^2-c)^{3/2}}
  \prod_{k=1}^n \frac{c}{(q_k^2-c)^{3/2}},\ \ n\geq 1.
\ee{W0p|q...}

In order to calculate the remaining correlators we first calculate
$W_0(\mid p,q)$. The bosonic part is twice the known universal
two--loop correlator of the complex matrix model whereas for the fermionic 
part we apply once more theorem \refer{theo} to eq.\ \refer{Wp0_0} and obtain
\bea
W_0(\,\mid p,q)&=&\frac{1}{2(p^2-q^2)^2}\biggl [ p^2\sqrt{\frac{q^2-c}{p^2-c}}
                 \ +\ q^2\sqrt{\frac{p^2-c}{q^2-c}}\ -\ 2pq \biggl ] 
\nonumber\\
&+&  \biggl ( \frac{1}{I_1}\dc\biggl )
\frac{c\, \X_2(\X_1-\L_0)}{I_1(p^2-c)^{3/2}(q^2-c)^{3/2}}.
\eea
The general result can be derived using the results of \cite{AJM} for the
bosonic part and theorem \refer{theo} for the fermionic part:
\be
W_0(\,\mid p_1,\ldots,p_n) =
\biggl [ \biggl ( \frac{1}{I_1}\dc\biggl )^{n-3}
   + \biggl ( \frac{1}{I_1}\dc\biggl )^{n-1}4\, \X_2\, (\X_1-\L_0)   \biggl ]
\frac{1}{4cI_1}\prod_{k=1}^n \frac{c}{(p_k^2-c)^{3/2}} 
\ee{W0(|...)}
for $n\geq 3$.

\sect{The Iterative Procedure}

\subsection{The Iteration for $u_1$ and $v_1$}

As already mentioned the structure of the superloop eqs.\ \refer{order0}
and \refer{order1} allow for an iterative solution in the genus, similar
in spirit to the situation for the supereigenvalue model \cite{Jan}.
Let us demonstrate how this works for $g=1$.
According to eq.\ \refer{order0} for $u_1(p)$ we have
\be
\widehat{\bar{V}}^\prime \, \circ\, u_1(p) = \frac{1}{2}\, \Bigl [
\, \dV{p}\, u_0(p) \, +\, \frac{1}{2\, p}\, \frac{d}{d\, q}\, \dY{p}\,
v_0(q)\, \mid_{p=q}\, \Bigr ] .
\ee{g=1o=0}
With the help of some formulas in the appendix, one shows that
\be
\dV{p}\, u_0(p) = \frac{1}{2\, p}\, \frac{d}{d\, q}\, \dY{p}\,
v_0(q)\, \mid_{p=q} \ = \frac{c^2}{8\, (p^2-c)^2\, p^2}.
\ee{du0}
And therefore
\be
u_1 (p) = -\frac{1}{8}\, \Y^{(1)} (p) + \frac{1}{8}\, \c^{(1)}(p)
+\frac{c}{8}\, \c^{(2)}(p) .
\ee{u1}
Now we solve the order 1 eq.\ \refer{order1} at $g=1$ for $v_1(p)$ 
\be 
\widehat{\bar{V}}^\prime \, \circ\, v_1(p) = 
\dV{p}\, v_0(p) \, - \, \widehat{\bar{\Y}} \, \circ\, u_1(p),
\ee{g=1o=1}
where we have introduced the operator \mat{\widehat{\bar{\Y}}}
defined by
\be
\widehat{\bar{\Y}} \, \circ \, f(p) = 
\cint{\w}{C}\, \frac{2\,\w^2 \,\bar{\Y} (\w )}{p^2-\w^2}\,
f(\w )\, -\, v_0(p)\, f(p).
\ee{defbarY}
Note that generally eq.\ \refer{g=1o=1} and its higher $g$ analogues fix
\mat{v_g(p)} only up to a zero--mode contribution \mat{\k_g\,\f^{(0)}(p)},
which will be determined later on. 
In order to evaluate the right hand side of eq.\ \refer{g=1o=1} we need
to know how the operator \mat{\widehat{\bar{\Y}}} acts on the basis functions
\mat{\f^{(n)}(p)} and \mat{\Omega^{(n)}(p)}. 
A straightforward calculation yields
\bea
\widehat{\bar{\Y}} \, \circ \, \f^{(n)} (p) &=&
\sum_{r=1}^{n+1}\, \frac{2\, \X_r}{(p^2-c)^{n+2-r}} 
-\frac{\X_1+\L_0}{(p^2-c)^{n+1}}\nonumber \\ &&
\nonumber \\
\widehat{\bar{\Y}} \, \circ \, \Omega^{(n)} (p) &=& 
\frac{\X_1-\L_0}{c^n}\, \frac{1}{p^2-c} \, -\, \sum_{r=1}^n
\sum_{l=1}^r\, \frac{2\, \L_{l-1}}{c^{n+1-r}}\, 
\frac{1}{p^{2(r+1-l)}}\nonumber\\
&& \sum_{r=1}^n \frac{\X_1+\L_0}{c^{n+1-r}}\, \frac{1}{p^{2r}}.
\eea
Moreover we need the quantity
\be
\delV{v_0(p)}{p}= W_0(p\mid p)= \frac{c\, (\L_0-\X_1 )}{2\,
I_1\,(p^2-c)^3}\,
+\, \frac{\L_0-\X_1+c\, \X_2}{2\, I_1\, (p^2-c)^2},
\ee{getW0pp}
which may be obtained from eq.\ \refer{W0p|q} by taking the limit $q\to p$.

We now have collected all the ingredients needed to evaluate the right 
hand side of eq.\ \refer{g=1o=1}. We find
\bea
v_1(p) &=& -\frac{\XL}{8\,c\,M_0}\, \Psi^{(1)}(p) - \frac{5\,c\XL}{8\, 
I_1}\, 
\chi^{(3)}(p)
\nonumber \\
&& - \Bigl [ \, \frac{3\,\XL}{8\, I_1}-\frac{c\, I_2\XL}{8\,{I_1}^2}
-\frac{c\,\X_2}{4\, I_1}\, \Bigr ]\, \chi^{(2)}(p)  
\\
&&- \Bigl [ \frac{\XL}{8\, c\, M_0}- \frac{\X_2}{4\, I_1}-\frac{c\, I_2\, 
\X_2}
{4\, {I_1}^2}+\frac{c\,\X_3}{4\, I_1}\, \Bigr ]\, \chi^{(1)}(p) 
+ \kappa_1\, \phi^{(0)}(p) \nonumber
\label{v1}
\eea
Yet the zero mode $\k_1$ is still undetermined.

\subsection{The Computation of $F_1$ and $\k_1$}

The bosonic part of the free energy of genus one $F_1^{\mbox{\scriptsize bos}}$
can be obtained by integrating eq.\ \refer{u1}:
\be
u_1(p)\ =\ \delV{F_1^{\mbox{\scriptsize bos}}}{p}.
\ee{intFbos}
The result is twice the free energy of the complex matrix model, when comparing
eq.\ \refer{u1} to \cite{AKM}, i.e.\
\be
F_1^{\mbox{\scriptsize bos}} \ =\ -\frac{1}{12}\ln I_1\ -\ \frac{1}{4}\ln M_0
\ -\ \frac{1}{3}\ln c \ .
\ee{Fbos}
In order to obtain the fermionic contribution to the free energy we rewrite
the known part of $v_1(p)$ in the following form:
\be
v_1(p)-\kappa_1\, \phi^{(0)}(p) =\frac{2}{c}\,\kappa_1\, \delY{\XL}{p}\,
 +
\, \delY{F_1}{p}.
\ee{inteq}
Upon using the relations
\bea
\f^{(0)}(p) &=& \frac{2}{c} \, \delY{\XL}{p} 
\nonumber\\
\f^{(k)}(p)&=& -\frac{2}{c}\, \dY{p}\, \Bigl[\, 
\sum_{l=2}^{k+1}\frac{\X_l}{
(-c)^{k-l+1}} +\frac{\XL}{(-c)^k}\, \Bigr] \\
\Omega^{(n)}(p)&=& 2\,\dY{p}\, \Bigl[\, 
\frac{\L_r}{c^{n+1-r}}-\frac{\XL}{c^{n+1}}\,
\Bigr ] \, ,\nonumber
\eea
one may integrate eq. \refer{inteq} to obtain the fermionic piece of the
free energy at genus one:
\bea
F_1^{\mbox{\scriptsize ferm}}&=& 
{{\X_2\,\X_3}\over {2\,{{I_1}^2}}} + 
  {\XL}\,\Bigl[ {{\L_1}\over {4\,{c^2}\,{{M_0}^2}}} - 
     \Bigl( {1\over {2\,{c^2}\,{{I_1}^2}}} + {{3\,I_2}\over 
{4\,c\,{{I_1}^3}}} 
\label{F_1^ferm} \\ && + 
        {{3\,{{I_2}^2}}\over {2\,{{I_1}^4}}} - {{5\,I_3}\over 
{4\,{{I_1}^3}}} + 
        {1\over {4\,{c^2}\,I_1\,M_0}} \Bigr) \,\X_2 + 
     {{\X_3}\over {2\,c\,{{I_1}^2}}} + {{3\,I_2\,\X_3}\over 
{2\,{{I_1}^3}}} - 
     {{5\,\X_4}\over {4\,{{I_1}^2}}} \Bigr].
\nonumber
\eea
Similarly the zero--mode $\kappa_1$ follows immediately:
\bea
\kappa_1&=& \XL \, \Bigl [ -\frac{5\, I_3}{8\,{I_1}^3}-\frac{1}{8\, 
c^2\, {M_0}^2}
+ \frac{1}{8\, c^2\, I_1\, M_0}+ \frac{3\, {I_2}^2}{4\, {I_1}^4}
\nonumber \\ && + \frac{3\, I_2}
{8\, c\, {I_1}^3} + \frac{1}{4\, c^2\, {I_1}^2}\, \Bigr ] + \frac{5\, c\, 
\X_4}
{8\, {I_1}^2} - \frac{\X_3}{2\, {I_1}^2}  -\frac{3\, c\, I_2\, \X_3}{4\, 
{I_1}^3}
\nonumber \\ &&
+ \X_2\, \Bigl ( \frac{3\, I_2}{8\, {I_1}^3} + \frac{3\, c\, {I_2}^2}{4\, 
{I_1}^4}
-\frac{5\, c\, I_3}{8\, {I_1}^3} + \frac{1}{8\, c\, I_1\, M_0} 
+ \frac{1}{4\, c\, {I_1}^2}\, \Bigr ) 
\nonumber \\ &&  -\frac{\L_1}{8\, c\, {M_0}^2}.
\eea

\subsection{The Computation of $\widehat{u}_1(p)$}

The remaining quantity $\widehat{u}_1(p)$ is now obtained from 
\refer{F_1^ferm}
by application of the $\delta/\delta V(p)$ operator. We find
\be
\widehat{u}_1(p)= \sum_{k=1}^4 A^1_k\, \chi^{(k)}(p)\, + B^1_1\,\Psi^{(1)}(p), 
\ee{u1p}
where
\bea
A^1_4 &=& {{-35\,c\,\XL\,\X_2}\over {8\,{{I_1}^2}}}
\nonumber \\ && \nonumber \\
A^1_3 &=& {{-15\,\XL\,\X_2}\over {8\,{{I_1}^2}}} + 
  {{25\,c\,\XL\,I_2\,\X_2}\over {8\,{{I_1}^3}}} - 
  {{15\,c\,\XL\,\X_3}\over {4\,{{I_1}^2}}}
\nonumber \\ && \nonumber \\
A^1_2 &=& {{3\,\XL\,I_2\,\X_2}\over {4\,{{I_1}^3}}} - 
  {{3\,c\,\XL\,{{I_2}^2}\,\X_2}\over {2\,{{I_1}^4}}} + 
  {{5\,c\,\XL\,I_3\,\X_2}\over {4\,{{I_1}^3}}} - 
  {{3\,\XL\,\X_2}\over {8\,c\,I_1\,M_0}}
\nonumber \\&&  - 
  {{3\,\XL\,\X_3}\over {2\,{{I_1}^2}}} + 
  {{3\,c\,\XL\,I_2\,\X_3}\over {{{I_1}^3}}} + 
  {{3\,c\,\X_2\,\X_3}\over {2\,{{I_1}^2}}} - 
  {{15\,c\,\XL\,\X_4}\over {4\,{{I_1}^2}}}
\nonumber \\ && \nonumber \\
A^1_1 &=& {{\X_2\, \L_1}\over {8\,c\,{{M_0}^2}}} - 
  {{\XL\,\X_2}\over {4\,{c^2}\,{{I_1}^2}}} - 
  {{3\,\XL\,I_2\,\X_2}\over {8\,c\,{{I_1}^3}}} - 
  {{3\,\XL\,{{I_2}^2}\,\X_2}\over {4\,{{I_1}^4}}}
\nonumber \\&& + 
  {{5\,\XL\,I_3\,\X_2}\over {8\,{{I_1}^3}}} + 
  {{9\,\XL\,I_2\,\X_3}\over {8\,{{I_1}^3}}} - 
  {{9\,c\,\XL\,{{I_2}^2}\,\X_3}\over {4\,{{I_1}^4}}} 
\nonumber \\&& - 
  {{3\,\XL\,\X_3}\over {8\,c\,I_1\,M_0}} + 
  {{5\,\X_2\,\X_3}\over {4\,{{I_1}^2}}} + 
  {{3\,c\,I_2\,\X_2\,\X_3}\over {4\,{{I_1}^3}}} - 
  {{15\,\XL\,\X_4}\over {8\,{{I_1}^2}}} 
\nonumber \\&&+ 
  {{15\,c\,\XL\,I_2\,\X_4}\over {4\,{{I_1}^3}}} + 
  {{5\,c\,\X_2\,\X_4}\over {8\,{{I_1}^2}}} - 
  {{35\,c\,\XL\,\X_5}\over {8\,{{I_1}^2}}}+ 
  {{15\,c\,\XL\,I_3\,\X_3}\over {8\,{{I_1}^3}}} 
\nonumber \\&&\nonumber \\
B^1_1 &=& -{{ \XL\,\L_1 }\over {4\,{c^2}\,{{M_0}^2}}} + 
  {{\XL\,\X_2}\over {8\,{c^2}\,I_1\,M_0}}.
\eea

This completes our computation of the genus one contributions
\mat{W_1(p|\, )}, \mat{W_1(\, | p)} and \mat{F_1}.

\sect{Conclusions}

We have constructed and completely solved a chiral supereigenvalue model 
away from the double scaling limit. The set of superloop equations
determining the correlation functions could be solved by an iterative
procedure in genus, in a similar way as for the supereigenvalue model 
\cite{Jan}. In addition all planar multi--superloop correlators were
determined explicitly, generalizing the results for the complex matrix model
\cite{AJM}.

It may be expected, that in the double scaling limit the model presented
here becomes equivalent to the double scaled supereigenvalue model
\cite{Alv2,Jan}, as the same relation holds for the complex and hermitian model
\cite{Amb}. This open question is left for further investigation.

\bigskip\noindent
\underline{Acknowledgments:} J.C.P. wishes to thank R. Mertig for advice with
{\it Mathematica}. The work of G.A. was supported by European Community
grant
no.\ ERBFMBICT960997.

\begin{appendix}
\pagebreak
\section*{Appendix}
\renewcommand{\theequation}{A.\arabic{equation}}
\setcounter{equation}{0}
Here we collect a number of important functional derivatives w.r.t. $V(p)$
\bea
\delV{c}{p} &=&  \frac{c}{I_1} \, \phi^{(1)}(p) \nonumber\\
\delV{I_k}{p} &=& (k+\frac{1}{2})\, \frac{c\, I_{k+1}}{I_1}\, 
\phi^{(1)}(p)
- (k+\frac{1}{2})\, c\, \phi^{(k+1)}(p) - k\, \phi^{(k)} \nonumber\\
\delV{M_k}{p} &=& (\frac{1}{2}-k)\, \Omega^{(k+1)}(p) 
+ \sum_{r=1}^k \frac{1}{2\, c^{k+1-r}}
\, \Omega^{(r)}(p) \nonumber\\&& 
- \sum_{r=0}^k \frac{M_r}{2\, c^{k-r}\, I_1}\, \phi^{(1)}(p) \nonumber\\
\delV{\L_k}{p} &=& \Bigl [ \,  \sum_{r=0}^k 
\frac{\L_r}{2\, c^{k-r}\, I_1} -\frac{\X_1}{2\, c^k\,  I_1}
\, \Bigr ]\, \phi^{(1)}(p) \nonumber\\
\delV{\X_k}{p} &=& \frac{(k-\frac{1}{2})\, c}{I_1}\, \X_{k+1}\, 
\phi^{(1)}(p)
\eea
and w.r.t. $\Y(p)$ using $\delY{\bar{\Y}(q)}{p}=\frac{q}{p^2-q^2}$:
\bea
\delY{\L_k}{p} &=& \delta_{k,0}\, \frac{p}{2} \, -\, \frac{1}{2}\, 
\frac{\sqrt{
p^2-c}}{p^{2k}} \nonumber\\
\delY{\X_k}{p} &=& \delta_{k,1}\, \frac{p}{2} \, -\, 
\frac{1}{2}\,\frac{p^2\,\sqrt{
p^2-c}}{(p^2-c)^k}.
\eea
The derivatives of the moments w.r.t. $c$ read
\bea
\delc{I_k}&=& (k+\frac{1}{2})I_{k+1} \nonumber\\
\delc{M_k}&=&-\sum^k_{r=0}\frac{M_r}{2c^{k-r+1}}+\frac{M_1}{2c^{k+1}}
\nonumber\\
\delc{\X_k}&=& (k-\frac{1}{2})\X_{k+1} \nonumber\\ 
\delc{\L_k} &=&\ \sum^k_{r=0}\frac{\L_r}{2c^{k-r+1}}-\frac{\L_1}{2c^{k+1}}.
\eea

\end{appendix}
\pagebreak

\end{document}